\documentclass[10pt,conference]{IEEEtran}
\ifCLASSOPTIONcompsoc
  \usepackage[nocompress]{cite}
\else
  \usepackage{cite}
\fi
\ifCLASSINFOpdf
\else
\fi
\usepackage{balance}  
\usepackage{graphics} 
\usepackage{txfonts}
\usepackage{times}    
\usepackage[pdftex]{hyperref}
\usepackage{color}
\usepackage{textcomp}
\usepackage{booktabs}
\usepackage{ccicons}

\usepackage{cite}
\usepackage{url}
\usepackage{fancybox}
\usepackage{multirow}
\usepackage{flushend}
\usepackage{booktabs}
\usepackage{tabularx}
\usepackage{comment}
\usepackage{array}
\usepackage[flushleft]{threeparttable}
\usepackage{mdframed}
\graphicspath{{}{images/}{dia/}}
\DeclareGraphicsExtensions{.pdf,.png}

\usepackage{listings}
\usepackage{courier}
\usepackage{hyperref}

\usepackage{xpatch}

\xpatchcmd{\refstepcounter}{%
  \stepcounter{#1}%
}{%
  \stepcounter{#1}%
  \xdef\scr{\number\value{#1}}%
}{\typeout{success}}{\typeout{failure}}

\newcounter{o}
\usepackage{tikz}
\usepackage{styles/pgf-pie}
\usetikzlibrary{positioning,shadows}
\usepackage{balance}

\newif\ifpienumberinlegend
\pgfkeys{/number in legend/.code=
    \expandafter\let\expandafter\ifpienumberinlegend
    \csname if#1\endcsname
    \ifpienumberinlegend

    \def\beforenumber##1\afternumber{}%
    \fi,
    /number in legend/.default=true
}
\definecolor{1c1}{RGB}{188,162,6}
\definecolor{1c2}{RGB}{137,129,80}
\definecolor{1c3}{RGB}{239,167,31}
\definecolor{1c4}{RGB}{88,194,241}
\definecolor{1c5}{RGB}{6,180,188}

\tikzset{mynode/.style={draw=white,solid,circle,fill=green,inner sep=1pt, thick,
text=black}}
\tikzset{arrow line/.style={dashed, line width= 2.5pt, color=#1}}

\def\bf{\textbf}

\def\fig {Fig.~}

\def\sec {Section~}
\def\secs {Sections~}

\def\it{\textit}

\def\tt{\mct}

\newcommand{\mct}[1]{{\footnotesize {\texttt {#1}}}}

\usepackage{paralist}

\newcommand{\nd}{\vspace{1mm}\noindent}

\usepackage{tikz}

\lstdefinestyle{inlinecode}{basicstyle={\ttfamily\scriptsize\bfseries}}

\newcommand{\urls}[1]{{\scriptsize\url{#1}}}
\usepackage{tcolorbox}

\usepackage{paralist}
\usepackage[outercaption]{sidecap}
\usepackage [autostyle, english = american]{csquotes}
\MakeOuterQuote{"}
\newcounter{scn}
\usepackage[shortlabels]{enumitem}
\usepackage{bchart}

\newcounter{finding_counter}
\newtoggle{comment}
\toggletrue{comment}

\begin{document}
\title{Automatic Detection of Five API Documentation Smells: Practitioners' Perspectives}

\author{
\IEEEauthorblockN{Junaed Younus Khan$^a$, Md. Tawkat Islam Khondaker$^a$, Gias Uddin$^b$ and Anindya Iqbal$^a$\\$^a$Bangladesh University of Engineering and Technology and $^b$University of Calgary}
}









\IEEEtitleabstractindextext{%
\begin{abstract}
The learning and usage of an API is supported by official documentation. 
Like source code, API documentation is itself a software product. 
Several research results show that bad design in API documentation 
can make the reuse of API features difficult. Indeed, similar to code smells or code anti-patterns, 
poorly designed API documentation can also exhibit 
`smells'. Such documentation smells can be described as bad documentation styles that
do not necessarily produce an incorrect documentation but nevertheless make the documentation difficult to
properly understand and to use. 
Recent research on API documentation has focused on finding content inaccuracies in API documentation and to complement 
API documentation with external resources (e.g., crowd-shared code examples).  
We are aware of no research that focused on the automatic detection of API 
documentation smells. This paper makes two contributions. First, we 
produce a catalog of five API documentation smells by consulting literature on API documentation presentation problems.
We create a benchmark dataset of 1,000 API documentation units by exhaustively and 
manually validating the presence
of the five smells in Java official API reference and instruction documentation. Second, we conduct a survey of 21 
professional software developers to validate the catalog. 
The developers agreed that they frequently encounter all five smells 
in API official documentation and 95.2\% of them reported that the presence of the  
documentation smells negatively affects their productivity. The participants wished for tool support to automatically detect and 
fix the smells in API official documentation. We
develop a suite of rule-based, deep and shallow machine learning classifiers to automatically detect the smells. 
The best performing classifier BERT, a deep learning model, achieves F1-scores of 0.75 - 0.97.  
\end{abstract}

\begin{IEEEkeywords}
API Documentation, Smell, Benchmark, Survey, Shallow Learning, Deep Learning.
\end{IEEEkeywords}}

%


\maketitle

\IEEEdisplaynontitleabstractindextext
\section{Introduction}\label{sec:introduction}
APIs (Application Programming Interfaces) are interfaces to reusable software
libraries and frameworks. Proper learning of APIs is paramount to support modern day rapid software
development. To support this, APIs typically are supported by official
documentation. An API documentation is a product itself, which warrants the
creation and maintenance principles similar to any existing software product. A
good documentation can facilitate the proper usage of an API, while a bad
documentation can severely harm its
adoption~\cite{Robillard-APIsHardtoLearn-IEEESoftware2009a,Robillard-FieldStudyAPILearningObstacles-SpringerEmpirical2011a,Aghajani-SoftwareDocPractitioner-ICSE2020}. 
 
A significant body of API documentation research has focused on studying API documentation problems based on surveys and interviews of software 
developers~\cite{Robillard-APIsHardtoLearn-IEEESoftware2009a,Robillard-FieldStudyAPILearningObstacles-SpringerEmpirical2011a,Aghajani-SoftwareDocPractitioner-ICSE2020,Cai-FrameworkDocumentation-PhDThesis2000,Carroll-MinimalManual-JournalHCI1987a,
Rossen-SmallTalkMinimalistInstruction-CHI1990a,Meij-AssessmentMinimalistApproachDocumentation-SIGDOC1992,Zhia-CostBenefitSoftwareDoc-JSS2015,Garousi-UsageUsefulnessSoftwareDoc-IST2015,Forward-RelevanceSoftwareDocumentationTools-DocEng2002}. 
Broadly, API documentation problems are divided into two types, what (i.e., what is documented) and 
how (i.e., how it is documented)~\cite{Aghajani-SoftwareDocIssueUnveiled-ICSE2019,Uddin-HowAPIDocumentationFails-IEEESW2015}. 
Tools and techniques are developed to address the `what' problems in API documentation, such as 
detection of code comment inconsistency~\cite{rabbi2020detecting,Wen-CodeCommentInconsistencyEmpirical-ICPC2019,Tan-tCommentCodeCommentInconsistency-ICSTVV2012,Zhou-DocumentationCodeToDetectDirectiveDefects-ICSE2017}, 
natural language summary generation of source code~\cite{McBurney-DocumentationSourceCodeSummarization-ICPC2014,Sridhara-SummaryCommentsJavaClasses-ASE2010,Haiduc:Summarization,Moreno-NLPJavaClasses-ICPC2013}, 
adding description of API methods by consulting external resources (e.g., online forums)~\cite{Aghajani-AndroidDocumentation-TSE2019}, 
detecting obsolete API documentation by comparing API version~\cite{Dagenais-DeveloperLearningResources-PhDThesis2012,Dagenais-TraceabilityLinksRecommendDocumentationEvolution-TSE2014},  
and complementing official documentation by incorporating insights and code examples from developer forums~\cite{Treude-APIInsight-ICSE2016,Subramanian-LiveAPIDocumentation-ICSE2014}. 
In contrast, 
not much research has focused on the automatic detection of `how' 
problems, e.g., bad design in API documentation that 
can make the reuse of API features difficult due to lack of usability~\cite{Uddin-HowAPIDocumentationFails-IEEESW2015}.
Recently, Treude et al.~\cite{Treude-DocumentationQuality-FSE2020} find that not all API documentation units are equally readable. This finding reinforces 
the needs to automatically detect API documentation presentation issues as `documentation smells', as previously highlighted 
by Aghajani et al.~\cite{Aghajani-SoftwareDocPractitioner-ICSE2020}.
Unfortunately, we are not
aware of any research on the automatic detection of such API documentation smells.

\begin{figure}[t]
\centering
	\centering
   	\includegraphics[scale=.7]{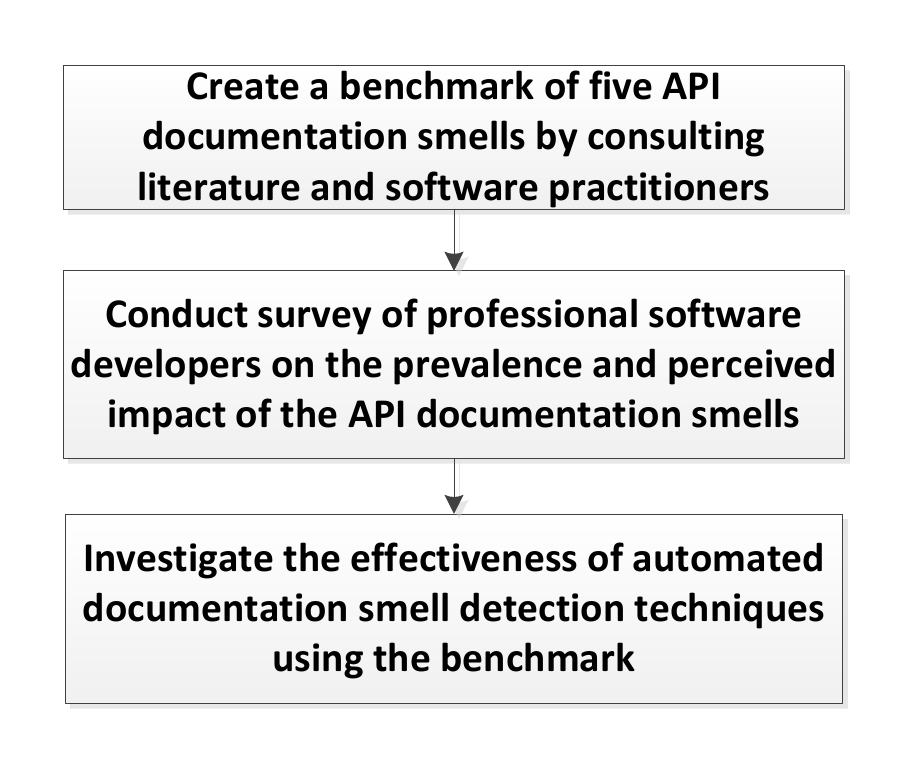}
   	\caption{The three major phases used in this study.}
   	 \label{fig:MethodologyOverall}
\vspace{-5mm}
\end{figure}
As a first step towards developing techniques to detect smells in API documentation, 
in this paper, we follow three phases (see \fig\ref{fig:MethodologyOverall}). First, we identify five API documentation 
smells by consulting API documentation literature~\cite{Uddin-HowAPIDocumentationFails-IEEESW2015,Aghajani-SoftwareDocIssueUnveiled-ICSE2019}  (\sec\ref{sec:benchmark}). 
Four of the smells (bloated, fragmented and tangled description of API documentation unit, and excess structural info in the description) are reported 
as presentation problems by Uddin and Robillard~\cite{Uddin-HowAPIDocumentationFails-IEEESW2015}. 
The other smell is called `Lazy documentation' and it 
refers to inadequate description of an API documentation 
unit (e.g., no explanation of method parameters). Such incomplete documentation is reported  
in literature~\cite{Aghajani-SoftwareDocIssueUnveiled-ICSE2019} and in online discussions. 
We exhaustively explore official API documentation to find the occurrences of the five smells. The focus  
was to develop a benchmark of \it{smelly} API documentation units. A total of 19 human coders 
participated in this exercise. This phase resulted in a benchmark of 1,000 API documentation units, where 778 units have at least one of the five smells.
{To the best of our knowledge, this is the first benchmark with real-world examples of the five documentation smells.}

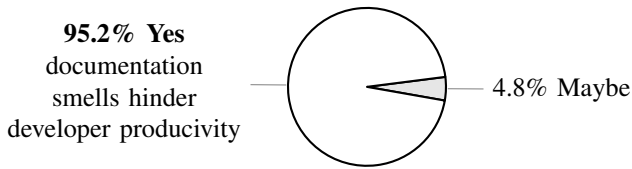
\begin{figure}[t]
	\centering\begin{tikzpicture}[scale=0.35]-
    \pie[
        /tikz/every pin/.style={align=center},
        text=pin, number in legend,
        explode=0.0,
        rotate = -10,
        color={black!10, black!0},
        ]
        {
            4.8/4.8\% Maybe,
            95.2/\bf{95.2\% Yes}\\ documentation\\ smells hinder\\developer producivity
        }
    \end{tikzpicture}
	\caption{Survey responses from professional developers on whether the presence of the smells in API documentation hinders productivity.}
	\vspace{-5mm}
	\label{fig:smell-productivity}
\end{figure}

In the second phase (\sec\ref{sec:survey}), we conducted a survey of 21 professional software developers to validate our catalog of API documentation smells. 
All the participants reported that they 
frequently encounter the five API documentation smells. More than 95\% of the 
participants (20 out of 21) reported that the presence of the five smells in API documentation negatively impacts their productivity (see \fig\ref{fig:smell-productivity}). 
The participants asked for tool support to automatically detect and fix the smells in API official documentation. These findings 
corroborate previous research 
that design and presentation issues in API documentation can hinder API usage~\cite{Uddin-HowAPIDocumentationFails-IEEESW2015,Aghajani-SoftwareDocPractitioner-ICSE2020}.
In the third phase (\sec\ref{sec:model}), we investigate a suite of rule-based, shallow and deep machine learning models using the benchmark to investigate 
the feasbility of automatically detecting the five smells. The best performing classifer BERT, a deep learning model, achieves F1-scores of 0.75 - 0.97.
{To the best of our knowledge, ours are the first techniques to automatically detect the five API documentation smells}.
{The machine learning models can be used to monitor and warn about API documentation 
quality by automatically detecting the smells in real-time with high accuracy.}

\nd\bf{Replication Package} with benchmark, code, and survey is shared at \url{https://github.com/disa-lab/SANER2021-DocSmell}
%
%
%

%

\section{A Benchmark of API Documentation Smells}\label{sec:benchmark}
We describe the methodology to create our benchmark of API documentation smells (\sec\ref{sec:benchmark_method}) and then present the 
benchmark with real-world examples (\secs\ref{sec:five_doc_smells} - \ref{sec:benchmark-details}).
\subsection{Benchmark Creation Methodology}\label{sec:benchmark_method}
\begin{figure}[t]
\centering
	\centering
   	\includegraphics[scale=.85]{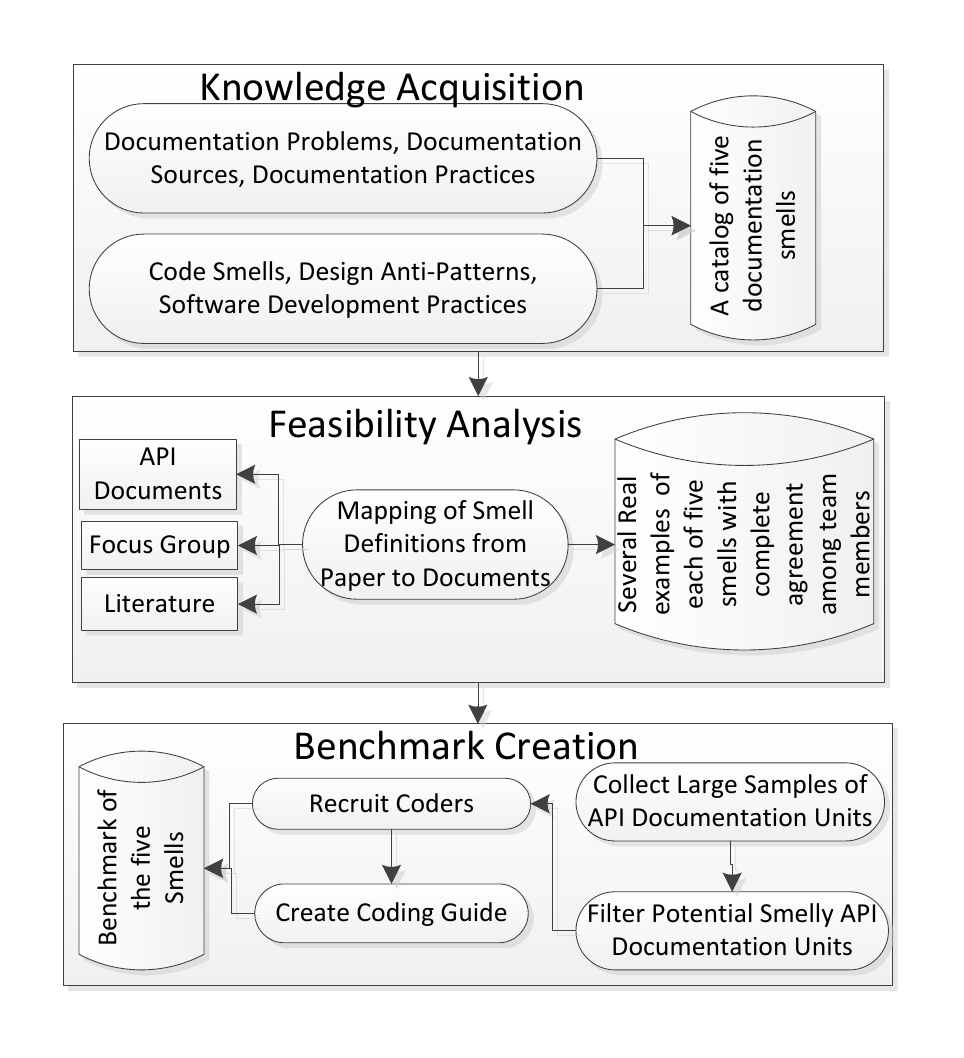}
   	\caption{The three major steps in benchmark creation process.}
   	 \label{fig:MethodologyOverallBenchmark}
\vspace{-5mm}
\end{figure}
Code and design smells are relatively well studied fields of software engineering. However, to the best of our knowledge, 
this is the first research on API documentation smells. As such, 
we needed to investigate both the literature 
on API documentation~\cite{Aghajani-SoftwareDocPractitioner-ICSE2020,Aghajani-SoftwareDocIssueUnveiled-ICSE2019,Uddin-HowAPIDocumentationFails-IEEESW2015,Robillard-FieldStudyAPILearningObstacles-SpringerEmpirical2011a} 
and the diverse API documentation resources (e.g., Java SE docs) during the creation of our catalog of API documentation smells. 
We followed a three-step process, which closely mimics the standard approaches followed in 
code/design smell formulation studies~\cite{Abidi-AntiPatternMultiLanguage-EuroPLoP2019,Abidi-CodeSmellsMultiLanguage-EuroPLoP2019}.  
The three steps are outlined in \fig\ref{fig:MethodologyOverallBenchmark} and are explained below.  

\begin{figure}[t]
  \centering
   \hspace*{-.7cm}%
  \includegraphics[scale=.16]{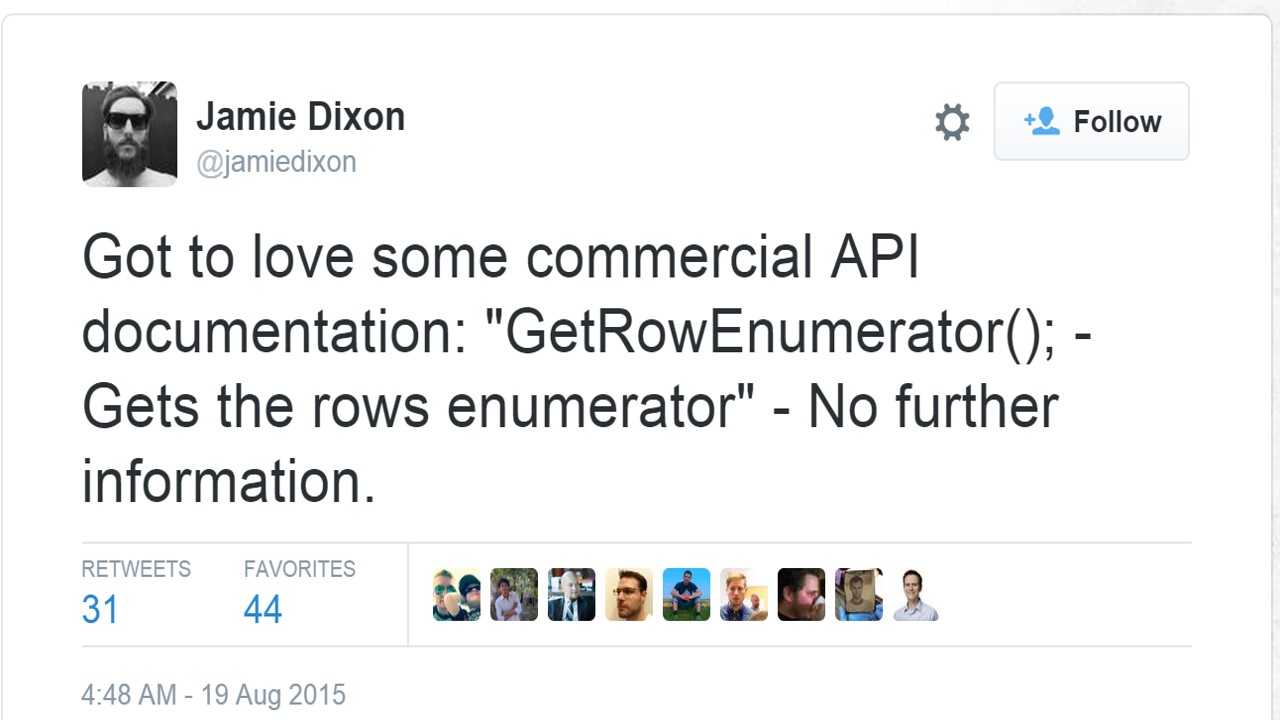}
  \caption{Tweet complaining about lazy documentation of API method.}

  \label{fig:motivating_lazy_example}
\vspace{-2mm}
\end{figure}

\nd\bf{Knowledge Acquisition.} Similar to code and design smells that do not directly introduce a defect or a bug into a software system, documentation smells refer to
presentation issues that do not make a documentation incorrect, rather they hinder its proper usage due to the lack of quality in the design 
of the documented contents. As such, we studied extensively the API documentation literature that reported issues related to API documentation presentation and usability~\cite{Uddin-HowAPIDocumentationFails-IEEESW2015,Aghajani-SoftwareDocIssueUnveiled-ICSE2019}.  
For example, the most recent paper on this topic was by Aghajani et al.~\cite{Aghajani-SoftwareDocIssueUnveiled-ICSE2019,Aghajani-SoftwareDocPractitioner-ICSE2020}, who 
divided the `how' problems in API documentation into four categories: maintainability (e.g., lengthy files), readability (e.g., clarity), 
usability (e.g., information organization like dependency structure), and usefulness (e.g., content not useful in practice). Previously, Uddin and Robillard~\cite{Uddin-HowAPIDocumentationFails-IEEESW2015} 
studied 10 common problems in API documentation by surveying 323 IBM developers. They observed four common problems related to presentation, i.e., bloated (i.e., too long description), 
tangled (complicated documentation), fragmented (i.e., scattered description), and excessive structural information (i.e., information organization like dependency structure).
Given that the four problems appeared in both studies, we included each as a documentation smell in our study. In addition, we added lack of 
proper description of an API method as a `lazy' documentation smell, because incomplete documentation problems are 
discussed in literature~\cite{Uddin-HowAPIDocumentationFails-IEEESW2015,Aghajani-SoftwareDocPractitioner-ICSE2020} as well as in online developer discussions  (see \fig\ref{fig:motivating_lazy_example}).  

\nd\bf{Feasibility Analysis.} Once we decided on the five smells, we conducted a feasibility study by looking for real-world examples of the smells in official
and instructional API documentation. This was important to ensure that the smells are prevalent in API documentation and that we can find those with reasonable confidence, because otherwise there is no way we can design automated techniques to detect those automatically. We combined our 
knowledge of the five smells gained from API documentation literature with active exploration of the five smells in the API official documentation. We conducted multiple 
focus group discussions where all the four authors discussed together by analyzing potential examples of the five smells in API documentation and by 
mapping the characteristics of such API documentation with the description of the smells in the literature/developer discussions. Before every such 
focus group meeting, the first two authors created a list of 50 API documentation units with their labels of the five smells in the units. The four authors 
discussed those labels together, refined the labels, and identified/filtered the labeling criteria. This iterative process led to increased understanding 
among the group members on the specific characteristics of the five documentation smells. From multiple discussion sessions, 
the final output was a list of 50 labeled datapoints.    

\nd\bf{Benchmark Creation.} In the last step of the benchmark creation process, we expanded our 
initial list of 50 API documentation units with smell labels as follows. 
We collected documentations of over 29K methods belonging to over 4K classes of
217 different packages. We extracted these documentations from the online JAVA
API Documentation
website~\cite{website:javadocse7} through web
crawling and text parsing techniques. 
Since a documentation can contain multiple smells at the
same time, this is a multi-labeled dataset. We produced the benchmark as
follows. First, all the authors mutually discussed the documentation smells.
Then, we randomly selected 950 documentations from a total of 29K that we
extracted. Then the first two authors labeled the first 50 documentations
separately. When they finished, they consulted other co-authors and resolved the
disagreement based on the discussion. Then they continued with the next 50
documentations and repeated the same process. Their agreement of labeling has been
recorded using Cohen's Kappa Coefficient \cite{cohen_kappa_paper} for each
iteration, i.e., labeling 50 documentations (Table \ref{cohen_kappa_table}).
After the third iteration, both the authors reached a perfect
agreement level with Cohen’s Kappa Coefficient of $0.83$. Then they prepared a
coding guideline for the labeling task which was later presented to 17 
computer science undergraduate students. The students labeled the remaining 800
documentation units. During the entire coding sessions by the 17 coders, the first 
two authors remained available to them via Skype/Slack. Each coder consulted their labels with the two authors. 
This ensured quality and mitigated subjective bias in the manual labeling of the benchmark. 


\begin{table}
\centering
\caption{Measure of agreement between two labelers}
\begin{tabular}{ccc}
\toprule
\textbf{Iteration ID}   &  \bf{Documentation Unit \#} & \bf{Cohen $\kappa$} \\ \midrule
 1 &  50                        & .49                   \\ 
 2 & 50               & .67                   \\ 
 3 & 50               & .83                   \\
\bottomrule
\end{tabular}
\label{cohen_kappa_table}
\end{table}

\subsection{The Five Documentation Smells in the Benchmark}
\label{sec:five_doc_smells}

\begin{figure}[h]
  \centering
  \includegraphics[scale=.48]{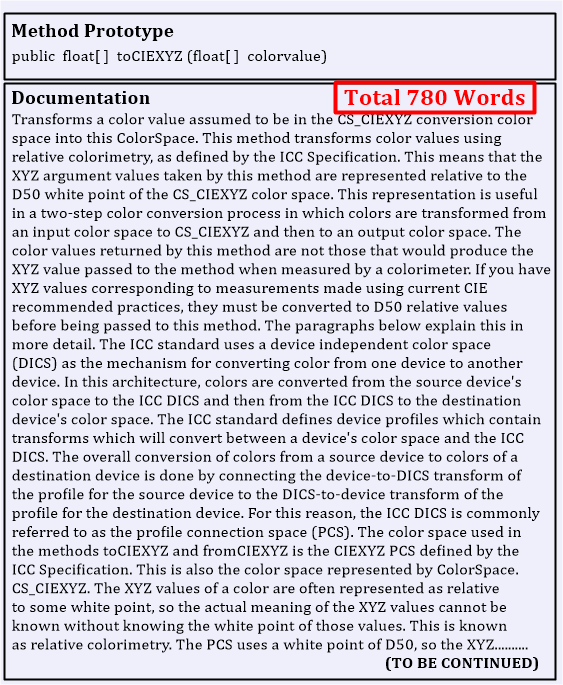}
  \caption{Example of Bloated Smell.}

  \label{fig:bloated_example}
\end{figure}

\nd\bf{Bloated Documentation Smell.} By `Bloated’ we mean the documentation 
whose description (of an API element type) is verbose or excessively elaborate. 
It is difficult to understand or follow a lengthy documentation~\cite{Uddin-HowAPIDocumentationFails-IEEESW2015}. 
Moreover, it cannot be effectively managed that makes it hard to modify when needed, e.g., in case of any update in the API source code. 
In our benchmark, we found many documentations that are larger than necessary. 
For example, the documentation shown in \fig\ref{fig:bloated_example} is so verbose and lengthy that it is hard to follow and use it. Hence, it is a bloated documentation.


\begin{figure}[h]
  \centering
  \includegraphics[scale=.48]{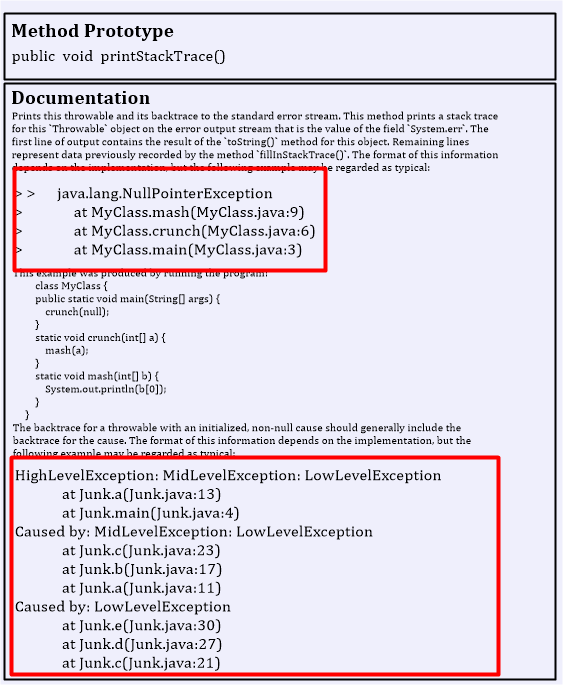}
  \caption{Example of Excess Structural Information.}

  \label{fig:excess_struct_example}
\end{figure}

\nd\bf{Excess Structural Information Smell.} Such a description of a documentation unit (e.g., method) 
contains too many structural syntax or information,
e.g., the Javadoc of the java.lang.Object class. Javadoc lists all the hundreds
of subclasses of the class. In our study, we find this type of documentation to
contain many class and package names. For instance, the documentation of \fig\ref{fig:excess_struct_example} contains many structural information (marked in
red rectangle) that are quite unnecessary for the purpose of understanding and
using the underlying method.


\begin{figure}[h]
  \centering
  \includegraphics[scale=.48]{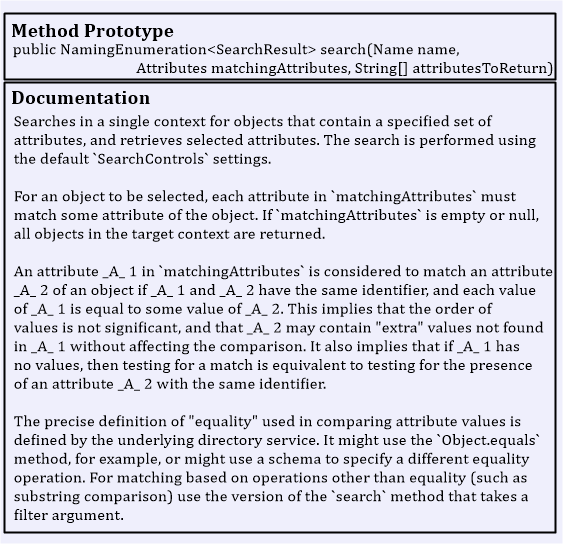}
  \caption{Example of Tangled Smell.}

  \label{fig:tangled_example}
\end{figure}

\nd\bf{Tangled Documentation Smell.} A documentation of an API element (method) is `Tangled’ if it's
description is \it{tangled} with various information (e.g., from other methods). This makes it complex and
thereby reduces the readability and understandability of the description. \fig\ref{fig:tangled_example} depicts an example of tangled documentation which is
hard to follow and understand.


\begin{figure}[h]
  \centering
  \includegraphics[scale=.48]{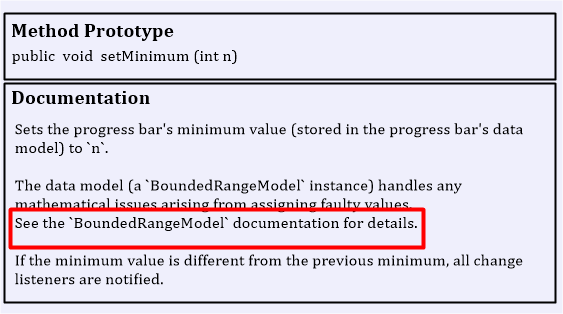}
  \caption{Example of Fragmented Smell.}

  \label{fig:fragmented_example}
\end{figure}

\nd\bf{Fragmented Documentation Smell.} Sometimes it is seen that the
information of documentation (related to an API element) is scattered (i.e., fragmented) over too
many pages or sections. In our
empirical study, we found a good number of documentation that contain many URLs
and references that indicate possible fragmentation smell. For example, the
documentation of \fig\ref{fig:fragmented_example} is fragmented as it refers
the readers to other pages or sections for details.

\begin{figure}[h]
  \centering
  \includegraphics[scale=.48]{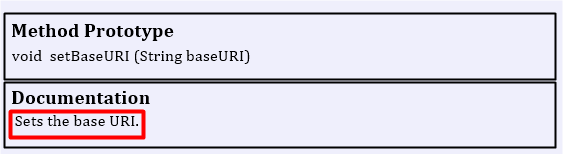}
  \caption{Example of Lazy Smell.}

  \label{fig:lazy_example}
\end{figure}

\nd\bf{Lazy Documentation Smell.} We categorize a documentation as `Lazy’ if it contains very small information to convey to the readers. In many cases, it is seen that the documentation does not contain any extra information except what can be perceived directly from the function name. Hence, this kind of documentation does not have much to offer to the readers. We see a lazy documentation in \fig\ref{fig:lazy_example} where the documentation says nothing more about the underlying method than what is suggested by the prototype itself. 


\subsection{Distribution of API Documentation Smells in Benchmark}\label{sec:benchmark-details}

\begin{figure}[t]
  \centering
  \includegraphics[scale=.19]{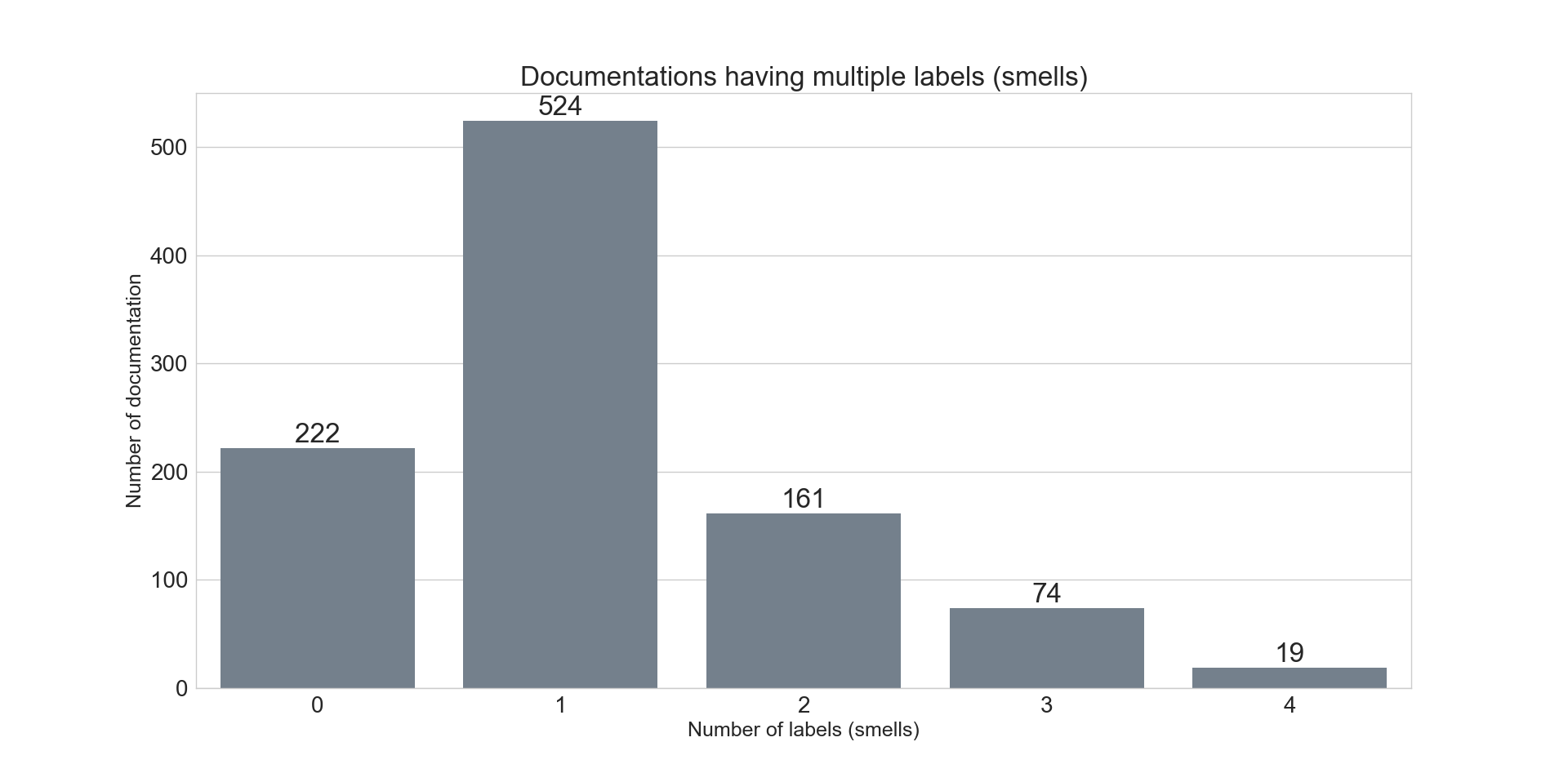}

  \caption{Smell distribution by \# of documentation units.}

  \label{fig:amount_of_multismell_in_dataset}

\end{figure}

\begin{figure}[t]
  \centering
   \includegraphics[scale=.19]{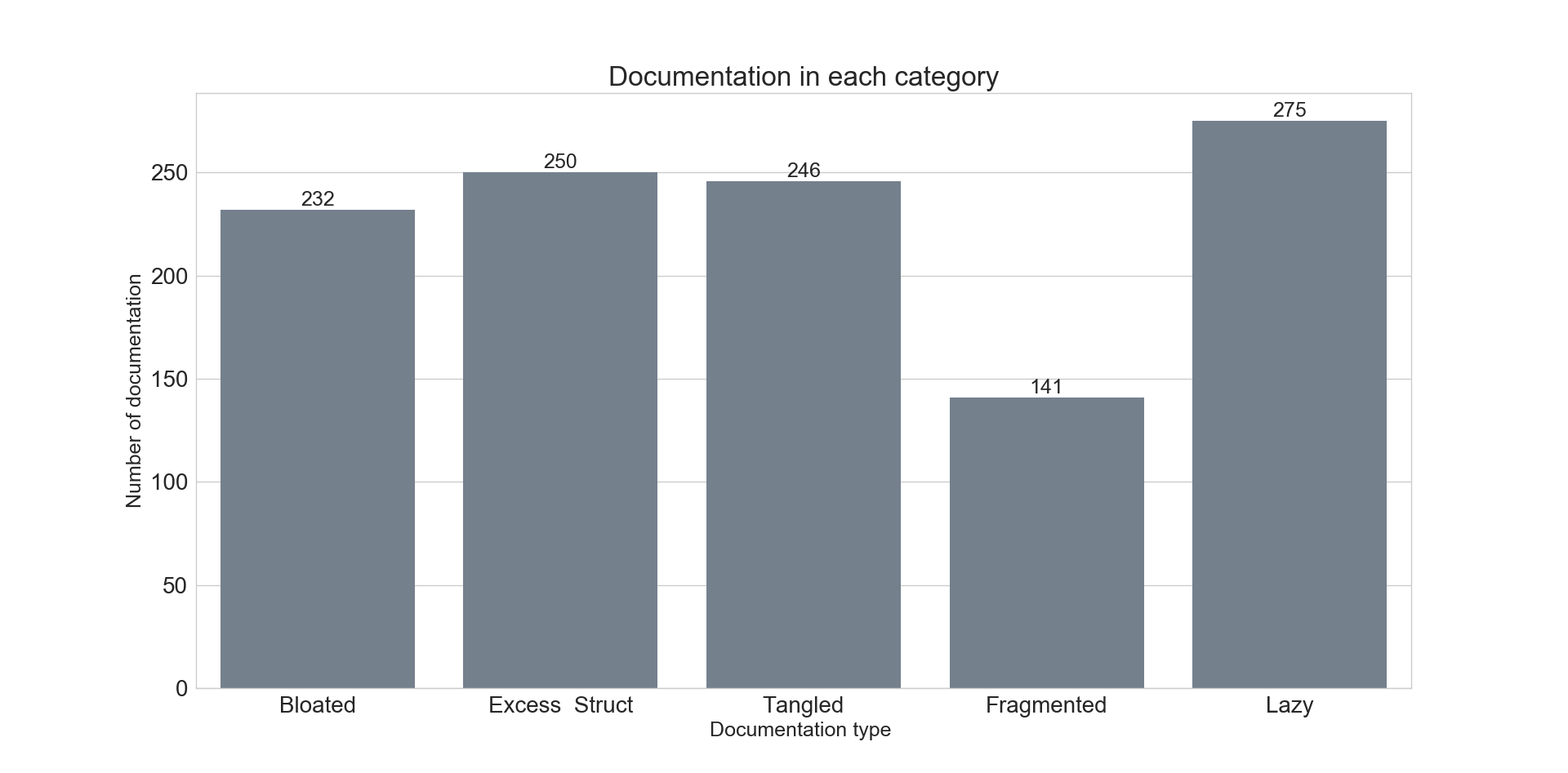}
  \caption{Distribution of different smells in our dataset.}

  \label{fig:distribution_of_smell_in_dataset}
\vspace{-4mm}
\end{figure}

We calculated the total number of smells in our dataset (\fig\ref{fig:amount_of_multismell_in_dataset}). 
We found that  778 documentations (almost 78\%) of our dataset contain at least one smell. While most (524) of the smelly documentations contain only one type of smell, a small number (19) of documentations show as high as four smells at the same time. We also determined the distribution of different smells in our dataset (\fig\ref{fig:distribution_of_smell_in_dataset}). It shows that all the five types of smells discussed occur in the dataset with a considerable frequency where the most frequent smell in our dataset is `Lazy’ with 275 occurrences and the least frequent smell is `Bloated’ with 141 occurrences.  


\begin{figure}[t]
  \centering
   \hspace*{-.7cm}%
  \includegraphics[scale=.5]{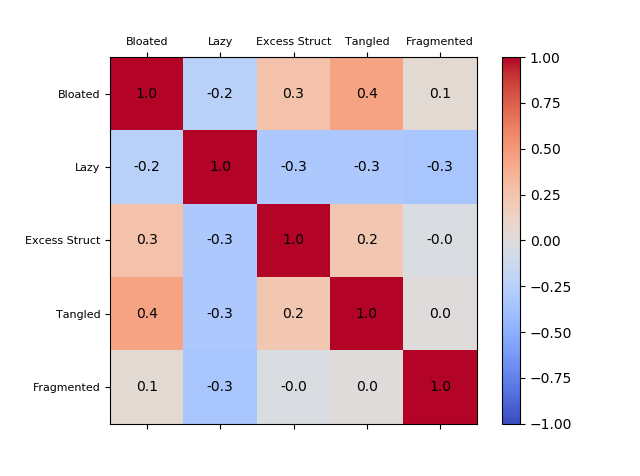}
  \caption{Correlation between different documentation smells in our benchmark. 
  Red, Blue, and Gray mean positive, negative, and no correlation. Intensity of color indicates the level of correlation.}

  \label{fig:label_correlation}
\vspace{-5mm}
\end{figure}

In multi-label learning, the labels might be interdependent and correlated
\cite{label_correlation_paper_1}. We used Phi Coefficients to determine such
interdependencies and correlations between different documentation smells. The
Phi Coefficient is a measure of association between two binary variables
\cite{phi_coefficient_paper}. It ranges from -1 to +1, where ±1 indicates a
perfect positive or negative correlation and 0 indicates no relationship. We
report the Phi Coefficients between each pair of labels in \fig\ref{fig:label_correlation}. We find that there is almost no correlation between
`Fragmented’ and any other smell (except `Lazy’). By definition, the information
of fragmented documentation is scattered in many sections or pages. Hence, it
has little to do with smells like `Bloated’, `Excess Structural Information’, or
`Tangled’. We also observe that there is a weak positive correlation (+0.2 to
+0.4) among the `Bloated’, `Excess Structural Information’, and `Tangled’
smells. One possible reason might be that if a documentation is filled with
complex and unorganized information (Tangled) or unnecessary structural
information (Excess Structural Information), it might be prone to become bloated
as well. On the other hand, `Lazy’ smell has a weak negative correlation (-0.2
to -0.3) with all other groups since these kinds of documentation are often too
small to contain other smells. However, none of these coefficients is high
enough to imply a strong or moderate correlation between any pair of labels.
Hence, all types of smells in our study are more or less unique in nature.

\section{Developers' Survey of Documentation Smells}\label{sec:survey}
Four out of the five API documentation smells in our study were previously reported as commonly observed by 
IBM developers~\cite{Uddin-HowAPIDocumentationFails-IEEESW2015}. 
The other smell (lazy documentation) is reported as a problem in API documentation in multiple studies~\cite{Aghajani-SoftwareDocPractitioner-ICSE2020,Robillard-APIsHardtoLearn-IEEESoftware2009a}. 
Given that we extended previous studies by creating a benchmark of the smells with real-world examples, we needed to further 
ensure that our collected examples of smelly documentation units do resonate with 
software developers. We, therefore, conducted a survey of professional software developers (1) to validate our catalog of the five API documentation smells and (2) to 
understand whether, similar to previous research, developers agree with the negative impact of the documentation smells. In particular, 
we explore the following two research questions:
\begin{enumerate}[label=RQ\arabic{*}., leftmargin=25pt]
  \item How do software developers agree with our catalog and examples of the five API documentation smells?
  \item How do software developers perceive the impact of the detected documentation smells?
\end{enumerate}\subsection{Survey Setup}
We recruited 21 professional software developers who are working in the software industry. 
We ensured that each developer is actively involved in daily software development activities like API reuse and documentation 
consultation. The participants were collected through personal contacts. First, each participant had to answer two demographic questions: current profession and years of experience in software development. 
We then presented each participant two Javadoc examples
of each smell and asked him/her whether they agreed that this documentation example
belonged to that particular smell. Then, we asked them about how frequently they
faced these documentation smells. Finally, we inquired them of the negative impact of the documentation smells on their overall productivity during software development. 
Out of the 21 participants, 14 participants had
experience less than 5 years and the rest had more than 5 years. Majority of the
participants had experience less than 5 years because they are likely to be more engaged in studying API documentation as part of their software programming 
responsibility. Developers with experience more than five years are more engaged in design of the software and its architecture.

\begin{figure}[t]
  \centering
  
  \includegraphics[scale=.4]{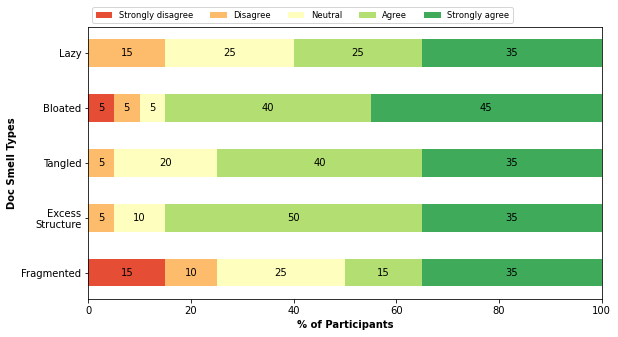}

  \caption{Survey response on whether the software developers agreed with our labeled documentation smell examples.}

  \label{fig:survey_examples}

\end{figure}
\begin{figure}[t]
  \centering
  
  \includegraphics[scale=.4]{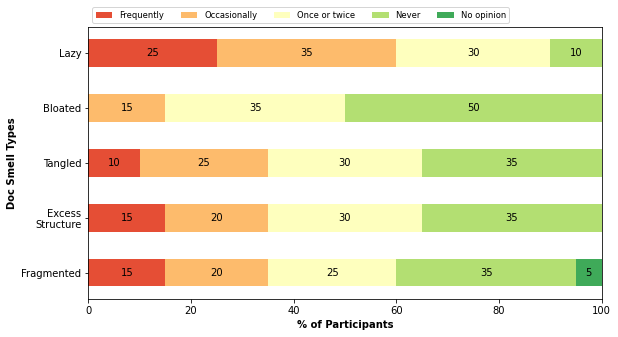}

  \caption{Survey response on how frequently the participants faced the documentation smells in the last three months.}

  \label{fig:survey_frequency}

\end{figure}

\subsection{How do software developers agree with our catalog and examples of the five API documentation smells? (RQ1)}
We showed each participant two examples of each smell, i.e., 10 examples in total. For each example, we asked 
two questions: (1) Do you think the documentation mentioned above is [smell, e.g., lazy]? The options are in Likert scale, i.e., strongly agree, agree, neutral, disagree, and strongly disagree. 
and (2) Based on your experience 
of the last three months, how frequently did you observe this [smell, e.g., lazy] in documentation? 
The options are: never, once or twice, occasionally, frequently, and no opinion. The options are picked from literature \cite{Uddin-HowAPIDocumentationFails-IEEESW2015}. 
Two examples per smell ensure increased 
confidence on the feedback we get from each participant.

\fig\ref{fig:survey_examples}
shows the responses of the participants to the first question. More than 75\% participants agreed to the examples of three smells: bloated, tangled, and 
excess structural info. At least 50\% of the participants agreed to the examples of the other two smells. Only 5-25\% of the participants 
disagreed to the examples. Overall, each example of the API documentation smell was agreed by at least 50\% of the participants. This validates 
out catalog of API documentation smells based on feedback from the professional developers. 

\fig\ref{fig:survey_frequency} shows the frequency of the documentation
smells the developers observed in the last three months (second question). We found that 50\% of
the participants had faced all the smells and lazy smell was the most
frequently encountered. On the other hand, half
of the participants did not face bloated documentation smells in the last three
months, while 60\%-65\% of the participants faced tangled, excess structural info,
and fragmented API documentation. This study reveals that API documentation is
becoming less explicable, more complex, and unnecessarily structured to keep the
documentation short. To solve this problem, API documentation
needs to be more understandable and elaborated to explain the API functionality.

\subsection{How do software developers perceive the impact of the detected documentation smells? (RQ2)}
We asked the participants how severely the documentation smells impact their
development tasks. The responses were taken on a scale of five degrees:
"Blocker", "Severe", "Moderate", "Not a Problem", and "No opinion". The options were picked from similar questions 
on API documentation presentation problems from literature~\cite{Uddin-HowAPIDocumentationFails-IEEESW2015}.
\begin{figure}[!htb]
  \centering
  
  \includegraphics[scale=.5]{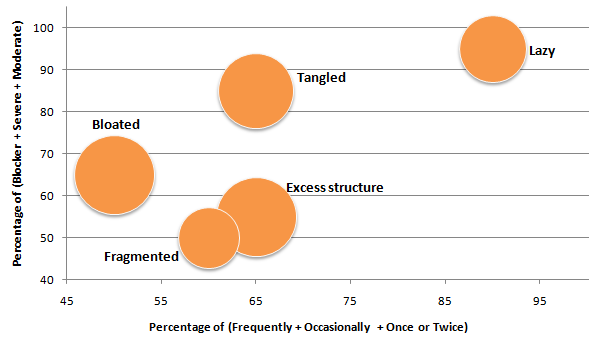}

  \caption{The perceived impact of the five documentation smells by severity and frequency. Circle size indicates the 
  percentage of participants who strongly agreed or agreed to the smells.}

  \label{fig:survey_severity}

\end{figure}

We analyzed the impact of the documentation smells with respect to the
frequency of the smells the participants had observed over the past three months (see \fig\ref{fig:survey_severity}). 
For each smell, we compute the frequency scale (x-axis) as the percentage of
response "Frequently", "Occasionally", and "Once or twice". For example, regarding whether the participants had
observed lazy documentation in the past three months, 25\% answered
"Frequently", 35\% answered "Occasionally", and 30\% answered "Once or twice",
leading to a total  90\% in the frequency scale. We constructed the severity
scale (y-axis) by combining the percentage of the participants responded with "Blocker",
"Severe", and "Moderate". For
example, due to fragmented documentation smells, 5\% of the participants could
not use that particular API and picked another API ("Blocker"), 20\% of the
participants believed that they wasted a lot of time figuring out the API
functionality ("Severe"), and 25\% of the participants felt irritated
("Moderate") with the fragmented documentation. The circle size indicates the
percentage of the participants "Strongly Agree" or "Agree" with the examples
containing documentation smells.

From \fig\ref{fig:survey_severity}, we observed that lazy documentation had
the most frequent and the most negative impact (90\%). Tangled documentation was
identified as the second most severe smell (85\%). Although bloated
documentation was considered more severe (65\%) than excess structural info (55\%
severity) and fragmented (50\% severity) documentation, bloated occurred less
frequently than the later two. The most important finding of this survey is that
the coordinates of all the circles (referring to documentation smells) in \fig\ref{fig:survey_severity} were above or equal to 50. This indicates that
according to the majority of the participants, these documentations smells are
occurring frequently and hindering the productivity of the development tasks.

\section{Automatic Detection of The Smells}\label{sec:model}
The responses from the survey validate our catalog of API documentation smells. 
The perceived negative impact of the smells on developers' productivity, as evidenced by the responses from our survey participants, necessitates the 
needs to fix API documentation by removing the smells. To do that, we first need to detect the smells automatically in the API documentation. The automatic detection offers 
two benefits: (1) we can use the techniques to automatically monitor and warn about bad documentation quality and (2) we can design techniques 
to fix the smells based on the detection. In addition, manual effort can also be made for improving detected examples. 
With a view to determine the feasibility of techniques to detect API documentation smells using our benchmark, 
we answer three research questions:
\begin{enumerate}[label=RQ\arabic{*}., start = 3, leftmargin=25pt]
  \item How accurate are rule-based classifiers to automatically detect the documentation smells?
  \item Can the shallow machine learning models outperform the rule-based classifiers?
  \item Can the deep machine learning models outperform the other models?
\end{enumerate}
The shallow and deep learning models are supervised, for which we used 5-fold
iterative stratified cross-validation as
recommended for a multilabel dataset in \cite{iterative_stratified_cross_valid}.
Traditional $k$-fold cross-validation is a statistical method of evaluating
machine learning algorithms which divides data into $k$ equally sized folds and
runs for $k$ iterations \cite{traditional_cross_validation_paper}. In each
iteration, each of the $k$ folds is used as the held-out set for validation
while the remaining $k-1$ folds are used as training sets. Stratified
cross-validation is used to make sure that each fold is an appropriate
representative of the original data by producing folds where the proportion of
different classes is maintained \cite{stratified_cross_validation_paper}.
However, stratification is not sufficient for multi-label classification
problems as the number of distinct labelsets (i.e., different combinations of
labels) is often quite large. For example, there can be 32 combinations of
labels in our study as there are 5 types of documentation smells. In such cases,
original stratified $k$-fold cross-validation is impractical since most groups
might consist of just a single example. Iterative stratification, proposed by
\cite{iterative_stratified_cross_valid}, solves this issue by employing a greedy
approach of selecting the rarest groups first and adding them to the smallest
folds while splitting. 

We report the performances using four standard metrics in information
retrieval~\cite{Manning-IRIntroBook-Cambridge2009}. Accuracy ($A$) is the ratio of
correctly predicted instances out of all the instances. Precision ($P$) is the ratio between the number of correctly predicted instances and all the predicted instances for a given smell.
Recall ($R$) represents the ratio of the number of correctly predicted instances
and all instances belonging to a given class. F1-score ($F1$) is the harmonic mean of
precision and recall.

{\scriptsize
\begin{eqnarray*}
P  = \frac{TP}{TP+FP},~
R = \frac{TP}{TP+FN},~
F1 = 2*\frac{P*R}{P+R},~
A = \frac{TP+TN}{TP+FP+TN+FN}
\end{eqnarray*}}
TP = Correctly classified as a smell, 
FP = Incorrectly classified as a smell, TN = Correctly classified as not a smell, 
FN = Incorrectly classified as not a smell.

\subsection{Performance of Rule-Based Classifiers (RQ3)}\label{sec:rule-based}
Based on manual analysis of a statistically significant random sample of our benchmark dataset (95\% confidence interval and 5 levels), 
we designed six metrics to establish five rule-based classifiers as described below. 
\subsubsection{Rule-based Metrics}
\label{subsubsec:rule-based-metrics}
\begin{inparaenum}[(a)]
\item\it{Documentation Length.} We use the length of every documentation in order to capture the extensiveness of the bloated documentations.
\item\it{Readability Metrics.} We measure Flesch readability metrics \cite{flesch_readability_paper} for the documentations 
to analyze the understandability of documentation. This feature might be useful to detect tangled documentations.
\item\it{Number of Acronyms and Jargons.} Since acronyms and jargons increase the
complexity of a reading passage \cite{jargon_paper_1}, we use the number of
acronyms and jargons in every documentation to detect the tangled documentation.
\item\it{Number of URLs} is computed because URLs are hints of possible fragmentation in the
documentation.
\item\it{Number of function, class, and package name mentioned} in documentation is computed to capture
excess structural information smell.
\item\it{Edit Distance.} The edit distance (i.e., measure of dissimilarity) between the
description of a lazy documentation and its' corresponding unit definition
(i.e., method prototype) can be smaller than non-lazy
documentations. We calculate the Levenshtein distance
\cite{edit_distance_levenshtein_paper} between the documentation description and
method prototype.
\end{inparaenum}

\subsubsection{Rule-based Classifiers}
\begin{figure}[t]
  \centering
  \hspace*{-.3cm}%
   \includegraphics[scale=.55]{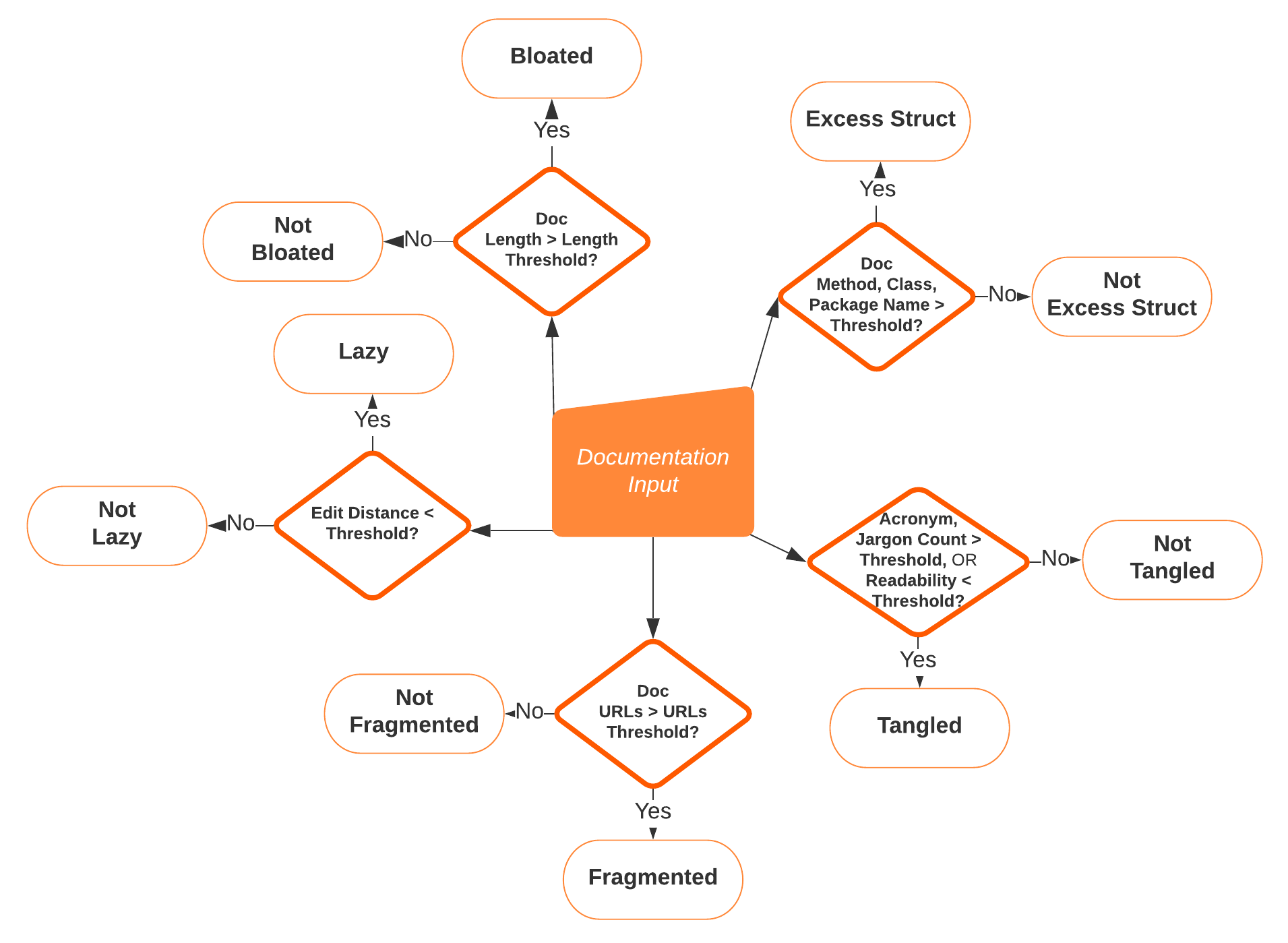}
  \caption{Flowchart of rule-based classification approach.}

  \label{fig:rule_based_classification_flowchart}
\vspace{-4mm}
\end{figure}

\fig\ref{fig:rule_based_classification_flowchart} shows flowchart of the rule-based classification approach.
For each metric, we study average, $25^{th}$, $50^{th}$, $75^{th}$, and $90^{th}$ percentiles as thresholds.

\subsubsection{Results}
\label{subsec:results_of_rulebased_baselines}


\begin{table*}[t]
\caption{Class-wise performance of rule-based baseline models by the metric thresholds (P stands for percentile)}
\begin{tabular}{p{.6cm}|lrrrr|rrrr|rrrr|rrrr|rrrr}
\multicolumn{1}{l}{} & \multicolumn{1}{l}{}                  & \multicolumn{4}{c}{\textbf{Bloated}}                                                                     & \multicolumn{4}{c}{\textbf{Lazy}}                                                                        & \multicolumn{4}{c}{\textbf{Excess Struct}}                                                             & \multicolumn{4}{c}{\textbf{Tangled}}                                                                     & \multicolumn{4}{c}{\textbf{Fragmented}}                                                                  \\
\midrule
{\textbf{Model}}   & \bf{Threshold}   & {A}   & {P}   & {R}   &{F1}   & {A}   & {P}   & {R}   &{F1}   & {A}   & {P}   & {R}   &{F1}   & {A}   & {P}   & {R}   &{F1}   & {A}   & {P}   & {R}   &{F1}  \\
\midrule
\multirow{5}{=}{\textbf{Rule Based}}
& {\textbf{AVG}}
& {.77} & {.38} & {.86} &{.52} 
& {.58} & {.39} & {.71} &{.51} 
& {.68} & {.35} & {.34} &{.34} 
& {.49} & {.13} & {.18} &{.15} 
& {.65} & {.33} & {.52} & {.40} 
\\ 
& {\textbf{25P}}
& {.39} & {.18} & {.64} &{.29} 
& {.96} & {.96} & {.93} & \bf{.95} 
& {.67} & {.38} & {.30} &{.34} 
& {.54} & {.09} & {.09} &{.09} 
& {.52} & {.31} & {.90} &\bf{.47} 
\\ 
& {\textbf{50P}}
& {.64} & {.28} & {.79} &{.41} 
& {.77} & {.55} & {.84} &{.66} 
& {.75} & {.37} & {.50} & \bf{.42} 
& {.45} & {.20} & {.40} &{.26} 
& {.61} & {.34} & {.71} &{.46} 
\\ 
& {\textbf{75P}}
& {.89} & {.56} & {.93} &{.70} 
& {.52} & {.36} & {.67} &{.47} 
& {.65} & {.32} & {.31} &{.31} 
& {.37} & {.25} & {.75} &{.37} 
& {.67} & {.29} & {.27} &{.28} 
\\ 
& {\textbf{90P}}
& {.95} & {.97} & {.85} & \bf{.90} 
& {.37} & {.30} & {.56} &{.39} 
& {.75} & {.50} & {.17} &{.25} 
& {.33} & {.26} & {.96} & \bf{.41} 
& {.72} & {.27} & {.10} &{.15} 
\\ 
\bottomrule
\end{tabular}
\label{table_individual_performance_baseline}
\end{table*}


In Table \ref{table_individual_performance_baseline}, we reported the
performances of the baseline models for each documentation smell. Different
thresholds of features achieved higher performance for different documentation
smells. For example, taking 90\textsuperscript{th} percentiles of the features’
values, baseline model achieved the higher performance for bloated documentation
detection, while lazy and excess structural information smell detection required
25\textsuperscript{th} percentile and 50\textsuperscript{th} percentile,
respectively. Notably, the performance of the baseline models in detecting
bloated (.90 F1-score) and lazy (F1 = .95) documentation were higher than
detecting excess structural info (F1 = .42), tangled (F1 = .41), and
fragmented (F1 = .47) documentations. 

\subsection{Performance of Shallow Learning Models (RQ4)}\label{sec:shallow-learning}
\subsubsection{Shallow Learning Models}
Since documentation smell detection is multi-label classification problem, we
employed different decomposition approaches: One-Vs-Rest (OVR), Label Powerset
(LPS), and Classifier Chains
(CC)~\cite{multilabel_decomp_1_MultilabelClassificationAnOverview,multilabel_decomp_2_AReviewOnMultilabelLearning,multilabel_decomp_3_ATutorialOnMultilabelClassification,multilabel_decomp_4_LearningFromMultilabelData}
with Support Vector Machine (SVM) \cite{SVM_SupportVectorNetworks} as the base
estimator. We chose SVM and OVR-SVM since those are successfully used for
multi-label text classification \cite{SVM_multi_1_AKernelMethodForMultilabelled,
SVM_multi_2_TextCategorizationWithSVM,svm_text_classification_1,
svm_text_classification_2,multilabel_decomp_1_MultilabelClassificationAnOverview}.
Each model trains a single classifier per class, with the samples of that class
as positive samples and all other samples as negatives. Each
individual classifier then separately gives predictions for unseen data.
We used linear kernel for the SVM classifiers as recommended by earlier works
\cite{svm_linear_kernel_paper_1, svm_linear_kernel_paper_2}.
\cite{svm_rbf_1_AComparisonStudyOfDifferentKernelFunctions,
svm_rbf_2_APracticalGuideToSVM}.   We also evaluated adapted approaches like
Multi label (ML) $k$NN \cite{MLkNN_ALazyLearningApproach} in this study. It
finds the $k$ nearest neighborhood of an input instance using $k$NN, then uses
Bayesian inference to determine the label set of the instance. We studied this
method because it has been reported to achieve considerable performance for
different multi-label classification tasks in previous studies
\cite{MLkNN_ALazyLearningApproach,
mlknn_MultilabelTextClassificationUsingSemanticFeatures}. For each algorithm, we
picked the best model using standard practices, e.g., hyper parameter tuning in
SVM as recommended by Hsu~\cite{Hsu-PracticalSVM-Misc2010}, choice of K in
ML-kNN as recommended by
\cite{mlknn_MultilabelTextClassificationUsingSemanticFeatures}.

\subsubsection{Studied Features}
We used two types of features: (1) rule-based metrics (described in Section
\ref{subsubsec:rule-based-metrics}) and (2) bag of words (BoW) \cite{bag_of_words_paper_1}. Bag of words (BoW) is a common feature extraction
procedure for text data and has been successfully used for text classification
problems \cite{bag_of_words_for_text_1,bag_of_words_for_text_2}.


\subsubsection{Results}

\begin{table*}[t]
\caption{Class-wise performance of shallow machine learning models}
\begin{tabular}{p{.75cm}|lrrrr|rrrr|rrrr|rrrr|rrrr}
\multicolumn{1}{l}{} & \multicolumn{1}{l}{}                  & \multicolumn{4}{c}{\textbf{Bloated}}                                                                     & \multicolumn{4}{c}{\textbf{Lazy}}                                                                        & \multicolumn{4}{c}{\textbf{Excess Struct}}                                                             & \multicolumn{4}{c}{\textbf{Tangled}}                                                                     & \multicolumn{4}{c}{\textbf{Fragmented}}                                                                  \\ 
\midrule
{\textbf{Feature}}   & {\textbf{Model}}   & {A}   & {P}   & {R}   &{F1}   & {A}   & {P}   & {R}   &{F1}   & {A}   & {P}   & {R}   &{F1}   & {A}   & {P}   & {R}   &{F1}   & {A}   & {P}   & {R}   &{F1}  \\ 
\midrule
\multirow{5}{=}{\textbf{Rule Based Feats}}  
& {\textbf{OVR-SVM}} 
& {.96} & {.88} & {.89} & \bf{.88} 
& {.94} & {.86} & {.94} &{.90} 
& {.74} & {.45} & {.23} &{.31} 
& {.82} & {.67} & {.56} &\bf{.61} 
& {.80} & {.69} & {.25} &{.37} 
\\ 
& {\textbf{LPS-SVM}} 
& {.94} & {.86} & {.70} &{.77} 
& {.91} & {.77} & {.97} &{.86} 
& {.74} & {.44} & {.21} &{.28} 
& {.80} & {.70} & {.40} &{.51} 
& {.81} & {.73} & {.32} &{.45} 
\\ 
& {\textbf{CC-SVM}}  
& {.96} & {.88} & {.87} &\bf{.88} 
& {.92} & {.79} & {.97} &{.87} 
& {.75} & {.47} & {.24} &{.32} 
& {.82} & {.68} & {.54} &{.60} 
& {.80} & {.71} & {.27} &{.39} 
\\ 
& {\textbf{ML-$k$NN}}  
& {.93} & {.73} & {.89} &{.80} 
& {.91} & {.86} & {.80} &{.83} 
& {.75} & {.49} & {.31} &{.38} 
& {.80} & {.63} & {.54} &{.58} 
& {.79} & {.57} & {.50} &{.53} 
\\ 
\midrule

\multirow{4}{=}{\textbf{BoW Feats}}
& {\textbf{OVR-SVM}}
& {.93} & {.84} & {.66} &{.74} 
& {.95} & {.87} & {.96} &\bf{.91} 
& {.75} & {.49} & {.47} &{.48} 
& {.78} & {.57} & {.54} &{.56} 
& {.79} & {.55} & {.54} &{.55} 
\\
& {\textbf{LPS-SVM}}
& {.93} & {.89} & {.63} &{.74} 
& {.94} & {.83} & {.97} &{.89} 
& {.75} & {.50} & {.49} &\bf{.50} 
& {.79} & {.59} & {.58} &{.58} 
& {.80} & {.59} & {.58} &\bf{.58} 
\\ 
& {\textbf{CC-SVM}}
& {.93} & {.85} & {.67} &{.75} 
& {.94} & {.85} & {.96} &{.90} 
& {.74} & {.48} & {.47} &{.48} 
& {.78} & {.57} & {.54} &{.56} 
& {.78} & {.54} & {.54} &{.54} 
\\ 
& {\textbf{ML-$k$NN}}
& {.93} & {.86} & {.60} &{.71} 
& {.88} & {.75} & {.83} &{.79} 
& {.73} & {.44} & {.29} &{.35} 
& {.79} & {.59} & {.53} &{.56} 
& {.80} & {.63} & {.41} &{.50} 
\\ 
\bottomrule
\end{tabular}
\label{table_individual_performance_shallow}
\end{table*}

Table \ref{table_individual_performance_shallow} presents the performance of the
shallow learning models. The best performer is OVR-SVM, followed closely by CC-SVM. CC-based
models are generally superior to OVR-based models because of the capability of
capturing label correlation \cite{CC_2_ClassifierChainsForMultilabel}. Since the
labels (types) of the presentation smells are not correlated (see 
\sec\ref{sec:benchmark-details}), the CC-based SVM could not
exhibit higher performance than the OVR-based SVM. Using rule-based features, OVR-SVM achieved a
higher F1-score (0.88) than the other models for bloated documentation
detection. Because documentation length (a rule-based) was more effective in
detecting bloated documentation than bag of words. On the other hand, LPS-SVM
achieved a higher F1-score (0.58) for fragmented documentation detection using
bag of words, as bag of words more successfully determined whether the
documentation was referring to other documentation than any rule-based features.
Overall, the shallow
models outperformed the rule-based classifiers for four smell types (except for
lazy documentation smell). Therefore, the documentation
smell detection does not normally depend on a single rule-based metric, rather, it depends
on a combination of different metrics and their thresholds. The shallow learning
models attempted to capture this combination of thresholds, and therefore,
achieved better performances than the baseline models.

\subsubsection{Feature Importance Analysis}
We verified the importance of our rule-based features by applying 
permutation feature importance technique \cite{permutation_feature_importance_paper_1,
permutation_feature_importance_paper_2} in the
best performing shallow model, i.e., OVR-SVM. We first train OVR-SVM with all
the features. While testing, we randomly shuffle the values
of one feature at a time while keeping other feature
values unchanged. A feature is important if shuffling its
values affects the model performance. We calculate the
change in performance in two ways. First, we measure the change in the
average F1-score of the OVR-SVM model for the permutation of a feature. Second,
we report the change of the specific class that the feature
was intended for (i.e., `Documentation  Length' for `Bloated'). We observe that
the permutation of any of our rule-based features degrades the model performance (see Table
\ref{table:permutation_feature_importance}). For example, after permutation of the
values of the `Documentation Length' of test data, the average F1-score
decreases by 0.17 (from 0.62 to 0.45) and the F1-score of the desired class
(i.e., `Bloated') decreases by 0.46 (from 0.88 to 0.42). This analysis
confirms the importance of combining rule-based metrics as features in the models.


\begin{table}
\centering
\caption{OVR-SVM performance decrease in feature permutation}
\begin{tabular}{llrr}
\toprule
\bf{Permuted} & \bf{Desired} & \multicolumn{2}{c}{\bf{Decrease in F1}}\\ \cline{3-4}
\bf{Feature} &  \bf{Class C} & \bf{Overall} & \bf{Desired C} \\ 
\midrule 
 Doc Length &  {Bloated}                      & {.17}          & {.46}            \\ 
 Readability & {Tangled}                      & {.06}          & {.11}                   \\ 
 \#Acronym\&Jargon & {Tangled}                & {.05}          & {.07}                   \\ 
 {\#URLs } & {Fragmented}                       & {.03}          & {.11}                   \\ 
 \#Method, Class, Package & {Excess Struct}     & {.17}          & {.08}                   \\ 
 Edit Distance & {Lazy}                       & {.09}          & {.37}                   \\
\bottomrule
\end{tabular}
\label{table:permutation_feature_importance}
\vspace{-.5cm}
\end{table}
\begin{table*}[t]
\caption{Class-wise performance of deep learning models}
\begin{tabular}{p{.75cm}|lrrrr|rrrr|rrrr|rrrr|rrrr}
\multicolumn{1}{l}{}&  \multicolumn{1}{l}{}                  & \multicolumn{4}{c}{\textbf{Bloated}}                                                                     & \multicolumn{4}{c}{\textbf{Lazy}}                                                                        & \multicolumn{4}{c}{\textbf{Excess Struct}}                                                             & \multicolumn{4}{c}{\textbf{Tangled}}                                                                     & \multicolumn{4}{c}{\textbf{Fragmented}}                                                                  \\ 
\midrule
{\textbf{Feature}}   & {\textbf{Model}}   & {A}   & {P}   & {R}   &{F1}   & {A}   & {P}   & {R}   &{F1}   & {A}   & {P}   & {R}   &{F1}   & {A}   & {P}   & {R}   &{F1}   & {A}   & {P}   & {R}   &{F1}  \\ 
\midrule
\multirow{2}{*}{\begin{tabular}[c]{@{}c@{}}\ \textbf{Word} \\   \textbf{Embed}\end{tabular}}
&{\textbf{Bi-LSTM}} & {.92} & {.92} & {.92} &{.91} & {.89} & {.90} & {.89} &{.90} & {.76} & {.72} & {{.76}} &{.73} & {.78} & {.74} & {.78} &{.74} & {.67} & {.64} & {.67} &{.63} \\ 
&{\textbf{BERT}}    & {.93} & {{.93}} & {{.93}} &{\textbf{.93}} & {{.97}} & {{.97}} & {{.97}} &{\textbf{.97}} & {.76} & {{.75}} & {{.76}} &{\textbf{.76}} & {.83} & {{.83}} & {{.83}} &{\textbf{.83}} & {.75} & {.75} & {{.75}} &{\textbf{.75}} \\ 
\bottomrule
\end{tabular}
\label{table_individual_performance_deep}
\end{table*}


\subsection{Performance of Deep Learning Models (RQ5)}\label{sec:deep-learning}
\subsubsection{Deep Learning Models}
We evaluated two deep learning models, Bidirectional LSTM (Bi-LSTM) and
Bidirectional Encoder Representations from Transformers (BERT).
We picked {Bi-LSTM}, because it is more capable of exploiting contextual information than the
unidirectional LSTM \cite{bilstm_FramewisePhonemeClassificationWithBiLSTM}.
Hence, the Bi-LSTM network can detect the documentation smell by capturing the
information of the API documentations from both directions. BERT is a  pre-trained model which was
designed to learn contextual word representations of unlabeled texts
\cite{bert_base_BertPretrainingOfDeepBidirectionalTransformers}. We picked BERT, because it is found to significantly outperform 
other models in various natural language processing and text
classification tasks
\cite{bert_success_1,bert_success_2,bert_success_3,bert_success_4,bert_success_5,bert_success_6,bert_success_7}.
We constructed a
Bi-LSTM model with 300 hidden states. We used ADAM optimizer
\cite{adam_optimizer} with an initial learning rate of 0.001. We trained the
model with batch size 256 over 10 epochs. 
We used BERT-Base for this study which has 12 layers with 12 attention heads and
110 million parameters. We trained it on benchmark for 10 epochs with
a mini-batch size of 32. We used early-stop to avoid overfitting
\cite{early_stop_bert_1} and considered validation loss as the metric of the
early-stopping \cite{early_stop_bert_2}. The maximum length of the input
sequence was set to 256. We used AdamW optimizer \cite{adam_w} with the learning
rate set to 4e\textsuperscript{-5}, $\beta1$ to 0.9, $\beta2$ to 0.999, and $\epsilon$ to
1e\textsuperscript{-8}
\cite{bert_base_BertPretrainingOfDeepBidirectionalTransformers,
bert_fine_tuning}. We used binary cross-entropy to calculate the loss
\cite{binary_cross_AreLossFunctionSame}.

\subsubsection{Studied Features}
We used word embedding as feature which
is a form of word representation that is capable of capturing the context of a
word in a document by mapping words with similar meaning to a similar
representation. For Bi-LSTM, we used 100-dimensional pre-trained GloVe embedding
which was trained on a dataset of one billion tokens (words) with a vocabulary
of four hundred thousand words \cite{Glove_GlobalVectorsForWordRepresentation}. We used the pre-trained embedding in BERT model \cite{bert_base_BertPretrainingOfDeepBidirectionalTransformers}.

\subsubsection{Results}

Table \ref{table_individual_performance_deep} shows the performance of the deep
learning models. BERT outperformed
Bi-LSTM, the shallow, and rule-based classifiers to detect each smell (F1-score). 
{The increase in F1-score in BERT compared to the best performing shallow learning model per smell is as follows: bloated (5.7\% over OVR-SVM Rule), 
lazy (6.6\% over OVR-SVM BoW), Excess Structural Information (52\% over ML-kNN BoW), tangled (36.1\% over OVR-SVM Rule), and 
fragmented (36.4\% over OVR-SVM BoW).} SVM and kNN-based models produced
more false-negative results because the number of positive instances for an
individual smell type is lower than the number of negative
instances for that type. As a result, SVM and kNN-based models showed low
recalls for some types (Excess structural information, Tangled, and Fragmented) and
consequently resulted in low F1-scores. On the other hand, Bi-LSTM and BERT achieved better performance because they
focused on capturing generalized attributes for each smell type. We manually analyzed the misclassified examples of Excess Structural Information and
fragmented documentation where BERT achieved below 0.8 accuracy. For the
Excess Structural Information smell detection, BERT falsely considered some java objects and
methods as structural information; therefore, the model produced some false
positive cases. In some examples, BERT could not identify whether the
information of documentation was referring to other documentation. As a result,
the model misclassified the fragmented documentation.

\section{Discussions}\label{sec:discussions}
\begin{figure}[t]
	\centering\begin{tikzpicture}[scale=0.25]-
    \pie[
        /tikz/every pin/.style={align=center},
        text=pin, number in legend,
        explode=0.0,
        rotate = -15,
        color={black!10, black!0},
        ]
        {
            9.5/9.5\% Maybe,
            90.5/\bf{90.5\% Yes}\\ smells should be\\ fixed in API\\ documentation
        }
    \end{tikzpicture}
	\caption{Survey responses on whether the five documentation smells should be fixed to improve API documentation quality.}
	\vspace{-5mm}
	\label{fig:smell-fix}
\end{figure}
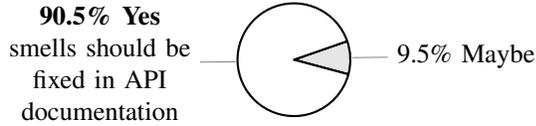

\nd\bf{Implications of Findings.} Thanks to the significant research efforts to understand API documentation problems using 
empirical and user studies, we now know with empirical evidence that the quality of API official documentation is a concern both for open source and 
industrial APIs~\cite{Uddin-HowAPIDocumentationFails-IEEESW2015,Robillard-FieldStudyAPILearningObstacles-SpringerEmpirical2011a,Aghajani-SoftwareDocIssueUnveiled-ICSE2019,Garousi-UsageUsefulnessSoftwareDoc-IST2015,Forward-RelevanceSoftwareDocumentationTools-DocEng2002}.  
The five API documentation smells we studied in this paper are 
frequently referred to as documentation presentation/design problems in the literature~\cite{Uddin-HowAPIDocumentationFails-IEEESW2015,Aghajani-SoftwareDocIssueUnveiled-ICSE2019}. 
Our comprehensive benchmark of 1,000 API official documentation 
units has 778 units each exhibiting one or more of the smells. The validity of the smells by professional software developers 
proves that this benchmark can be used to foster a new area of research in software engineering on the automatic detection 
of API documentation quality - which is now an absolute must due to the growing importance of APIs and software in our daily lives~\cite{Robillard-FieldStudyAPILearningObstacles-SpringerEmpirical2011a,Ponzanelli-PrompterRecommender-EMSE2014}. 
The superior performance of our machine learning classifiers, in particular the deep learning model BERT, offers promise that we can now use 
such tools to automatically monitor and warn about API documentation quality in real-time.  
Software companies and open source community can leverage our developed model to analyze the quality of their API documentation. 
Software developers could save time by focusing on good quality API documentation instead of the bad ones as detected by our model.
Based on such real-time feedback, tools 
can be developed to improve the documentation quality by fixing the smells. Indeed, when we asked our survey participants (\sec\ref{sec:survey}) whether 
the five smells need to be fixed, more than 90\% responded with a `Yes', 9.5\% with a `Maybe', 0\% with a `No' (see \fig\ref{fig:smell-fix}).

\nd\bf{Threats to Validity.}  \it{Internal validity} threats relate to authors' bias 
while conducting the analysis. We mitigated the bias in our benchmark creation process by taking agreement 
from 17 coders and co-authors and by consulting API documentation literature. The machine learning models 
are trained, tested, and reported using standard practices. There was no common data between the training and test set. 
\it{Construct validity} threats relate to the difficulty in finding
data to create our catalog of smells. Our benchmark 
creation process was exhaustive, as we processed more than 29K unit examples from official documentation.  
\it{External validity} threats relate to the
generalizability of our findings. We mitigated this threat by corroborating the five smells in our study 
with findings from state-of-the-art research in API documentation presentation and design problems. 
Our analysis focused on the validation and detection of five API documentation smells. Similar to code smell literature, 
additional documentation smells can be added into our catalog as we continue to research on this area.

\section{Related Work}\label{sec:related-work}
Related work is divided into \bf{studies} on understanding (1) documentation problems and (2) how developers learn APIs using documentation, and developing 
\bf{techniques} (3) to detect errors in documentation and (4) to create documentation.

\nd\bf{Studies.} Research shows that traditional Javadoc-type approaches to 
API official documentation are less useful than example-based documentation (e.g., minimal manual~\cite{Carroll-MinimalManual-JournalHCI1987a})~\cite{Shull-InvestigatingReadingTechniquesForOOFramework-TSE2000}
Both code examples and textual description are required for better quality API documentation~\cite{Forward-RelevanceSoftwareDocumentationTools-DocEng2002,DeSouza-DocumentationEssentialForSoftwareMaintenance-SIGDOC2005,Nykaza-ProgrammersNeedsAssessmentSDKDoc-SIGDOC2002}. 
Depending of the types of API documentation, reability and understandability of the documentation can vary~\cite{Treude-DocumentationQuality-FSE2020}. 
Broadly, problems in API official documentation can be about `what' contents are documented and `how' the contents are presented~\cite{Uddin-HowAPIDocumentationFails-IEEESW2015,Aghajani-SoftwareDocIssueUnveiled-ICSE2019,Robillard-APIsHardtoLearn-IEEESoftware2009a,Robillard-FieldStudyAPILearningObstacles-SpringerEmpirical2011a,Aghajani-SoftwareDocPractitioner-ICSE2020}. 
Literature in API documentation quality discussed four desired attributes of API documentation: 
completeness, consistency, usability and accessibility \cite{Zhia-CostBenefitSoftwareDoc-JSS2015,Treude-DocumentationQuality-FSE2020}. 
Several studies show that external informal resources can be consulted to improve API official 
documentation~\cite{Yang-QueryToUsableCode-MSR2016,Kavaler-APIsUsedinAndroidMarket-SOCINFO2013,Wang-APIsUsageObstacles-MSR2013,Sunshine-APIProtocolUsability-ICPC2015,Parnin-MeasuringAPIDocumentationWeb-Web2SE2011,Delfim-RedocummentingAPIsCrowdKnowledge-JournalBrazilian2016,Jiau-FacingInequalityCrowdSourcedDocumentation-SENOTE2012}

The five documentation smells studied in this paper are taken from five commonly discussed API 
documentation design and presentation issues in literature~\cite{Uddin-HowAPIDocumentationFails-IEEESW2015,Aghajani-SoftwareDocPractitioner-ICSE2020}. 
In contrast to the above papers that aim to understand API documentation problems, we focus on the development of techniques to automatically detect documentation smells.  

\nd\bf{Techniques.} Tools and techniques are proposed to automatically add code examples and insights from external resources (e.g., online forums) into 
API official documentation~\cite{Subramanian-LiveAPIDocumentation-ICSE2014,Treude-APIInsight-ICSE2016,Aghajani-AndroidDocumentation-TSE2019}. Topic modeling 
is used to develop code books and to detect deficient documentation~\cite{Souza-CookbookAPI-BSSE2014,Souza-BootstrapAPICodeBookSO-IST2019,Campbell-DeficientDocumentationDetection-MSR2013}. 
API official documentation and online forum data are analyzed together to recommend fixes API misuse scenarios~\cite{Ren-DemystifyOfficialAPIUsageDirectivesWithCorwdExample-ICSE2020}.  
The documentation of an API method can become obsolete/inconsistent due to evolution in source code~\cite{Wen-CodeCommentInconsistencyEmpirical-ICPC2019,Dagenais-DeveloperLearningResources-PhDThesis2012}. 
Several techniques are proposed to automatically detect code comment inconsistency~\cite{rabbi2020detecting,Tan-tCommentCodeCommentInconsistency-ICSTVV2012,Zhou-DocumentationCodeToDetectDirectiveDefects-ICSE2017}. 
A large body of research is devoted to automatically produce natural lanugage summary description of source code method~\cite{McBurney-DocumentationSourceCodeSummarization-ICPC2014,Sridhara-SummaryCommentsJavaClasses-ASE2010,Haiduc:Summarization,Moreno-NLPJavaClasses-ICPC2013}. 

Unlike previous research, we focus on the detection of five API documentation smells that do not make a 
documentation inconsistent/incorrect, but nevertheless make the learning of the documentation difficult due to the underlying design/presentation issues. 
We advance state-of-the-art research on API documentation quality analysis by offering a benchmark of real-world examples of five documentation smells 
and a suite of techniques to automatically detect the smells. 
\section{Conclusions}
The learning of an API is challenging when the official documentation resources are of low quality. We identify 
five API documentation smells by consulting API documentation literature on API documentation design and presentation issues.  
We present a benchmark of 1,000 API documentation units with five smells in API official documentation. 
Feedback from 21 industrial software developers shows that the smells can negatively impact the productivity of the developers 
during API documentation usage. We
develop a suite of machine learning classifiers to automatically detect the smells. 
The best performing classifier BERT, a deep learning model, achieves F1-scores of 0.75 - 0.97. 
The techniques can help automatically monitor and warn about API documentation quality.

%

\begin{small}
\bibliographystyle{abbrv}
\bibliography{consolidated}
\end{small}

\end{document}